\documentclass[showpacs,twocolumn,amssymb,prl,aps]{revtex4}
\usepackage{graphicx,epsfig}

\newcommand{\dC}{$^{\circ}$C}

\newcommand{\etal}{\textit{et. al.}}

\begin{document}
\title{Continuous Paranematic-to-Nematic Ordering Transitions of\\ Liquid Crystals in Tubular Silica Nanochannels}
\author{Andriy~V.~Kityk$^{1}$ }
\email[E-mail: ]{andriy.kityk@univie.ac.at}
\author{Matthias~Wolff$^2$}
\author{Klaus~Knorr$^2$}
\author{Denis~Morineau$^3$}
\author{Ronan Lefort$^3$}
\author{Patrick~Huber$^2$}
\email[E-mail: ]{p.huber@physik.uni-saarland.de}
\affiliation{$^{1}$Institute for Computer Science, Czestochowa University of Technology, Al. Armii Krajowej 17, P-42200 Czestochowa, Poland\\ $^2$Faculty of Physics and Mechatronics Engineering, Saarland University, D-66041 Saarbr\"ucken, Germany\\ $^3$Institut de Physique de Rennes, CNRS-UMR 6251, Universit\'{e} de Rennes 1, F-35042 Rennes, France}

\date{\today}

\begin{abstract}
The optical birefringence of rod-like nematogens (7CB, 8CB), imbibed in parallel silica channels with 10~nm diameter and 300 micrometer length, is measured and compared to the thermotropic bulk behavior. The orientational order of the confined liquid crystals, quantified by the uniaxial nematic ordering parameter, evolves continuously between paranematic and nematic states, in contrast to the discontinuous isotropic-to-nematic bulk phase transitions. A Landau-de~Gennes model reveals that the strength of the orientational ordering fields, imposed by the silica walls, is beyond a critical threshold, that separates discontinuous from continuous paranematic-to-nematic behavior. Quenched disorder effects, attributable to wall irregularities, leave the transition temperatures affected only marginally, despite the strong ordering fields in the channels.
\end{abstract}

\maketitle

Spatial confinement on the micro- and nano-scale can affect the physics of liquid crystals (LCs) markedly. Mo{\-}dified phase transition behavior has been found in ex{\-}periments on LCs imbibed into a variety of porous media \cite{I1, Kralj}, in aerogels \cite{I2}, in semi-confined thin film geometries \cite{Garcia2008}, and at the free surface of bulk LCs \cite{Ocko1986}.

For example, the heat capacity anomaly typical of the second-order nematic-to-smectic-A (\textit{N-SmA}) transition in rod-like LCs immersed in aerogels is absent or greatly broadened. This allowed a detailed study of the influence of quenched disorder introduced by random spatial confinement on this archetypical phase transition \cite{I2}.

It has also been demonstrated experimentally \cite{I1, Kralj}, in agreement with expectations from theory \cite{Sheng1976, Steuer2005}, that there is no ''true'' isotropic-nematic \textit{(I-N)} transition for LCs confined in geometries spatially restricted in at least one direction to a few nanometers. The anchoring at the confining walls, quantified by a surface field, imposes a partial orientational, that is a partially nematic ordering of the confined LCs, even at temperatures $T$ far above the bulk \textit{I-N} transition temperature $T^{\rm b}_{IN}$. The symmetry breaking doesn't occur spontaneously, as characteristic of a genuine phase transition, but is enforced over relevant distances by the interaction with the walls \cite{Stark2002}. Thus confinement here plays a similar role as an external magnetic field for a spin system: The strong first order \textit{I-N} transition is replaced by a weak first order or continuous paranematic-to-nematic (\textit{P-N}) transition, depending on the strength of the surface orientational field.

Whereas a qualitative understanding of this behavior has been achieved for a variety of spatially mesoconfined LCs \cite{I1, Kralj}, more detailed comparisons with theoretical predictions have been extremely challenging in the past, mainly due to the complex, tortuous, multiply connected pore networks or unknown surface/interface-LC interactions in the aforementioned studies. The advent of arrays of straight nanochannels of silica and silicon with simpler channel geometries may allow to gain deeper, quantitative insights into this phenomenology.

In this Letter, we present a high-resolution optical birefringence study of rod-like nematogens (7CB and 8CB) confined to an array of parallel, non-tortuous channels of 10~nm mean diameter and 300~micrometer length in a monolithic silica membrane. We demonstrate that the optical transparency, the straight channel geometry of the host along with the sensitivity of the modulated beam technique employed allows us to precisely characterize the orientational order of the nematogens with respect to the long axis of the nanochannels. The simple type of restricted geometry will allow us to compare our measurements quantitatively with a Landau-de Gennes model for the \textit{P-N} transition and, consequently, to determine both the strength of the orientational ordering field, imposed by the silica walls, and the influence of quenched disorder, attributable to channel irregularities.

\begin{figure*}[htbp]
\epsfig{file=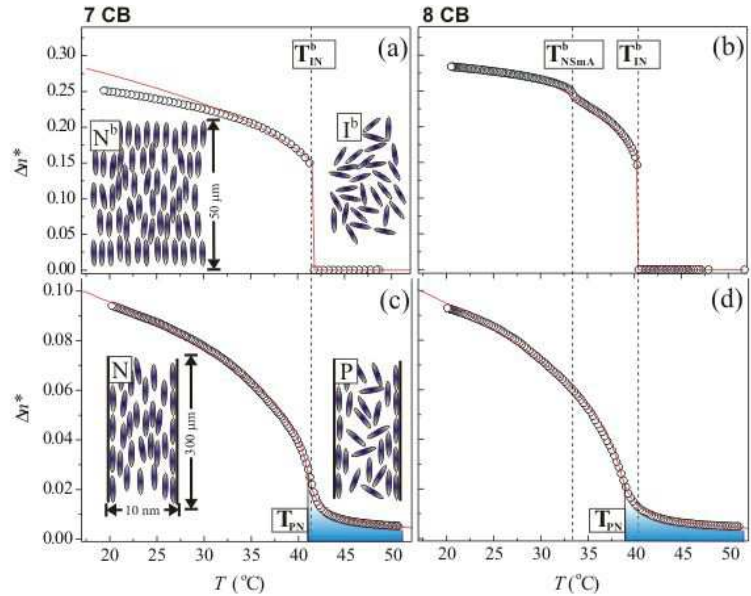, angle=0,
width=1.2\columnwidth}\caption{(color online). Birefringence of 7CB and 8CB measured in the bulk state, panel (a), (b), and in the silica nanochannels, panel (c) and (d), resp., as a function of temperature in comparison to fits (solid lines) based on the KKLZ-model discussed in the text. The final birefringence characteristic of the paranematic phases are shaded down to the $P-N$ ''transition'' temperatures, $T_{\rm PN}$. The dashed lines mark the bulk \textit{I-N} and \textit{N-SmA} transition temperatures. As insets in (a) and (c), the bulk isotropic ($I^{\rm b}$) as well as the bulk nematic ($N^{\rm b}$) phases upon homeotropic alignment, and the confined paranematic (P) and nematic (N) phases are illustrated, respectively.}
\label{fig1}
\end{figure*}

For rod-like molecules the degree of orientational molecular order can be quantified by the uniaxial order para{\-}meter $Q= \frac{1}{2}\left \langle 3 \cos^2\phi -1 \right \rangle$, where $\phi$ is the angle between the long axis of a single molecule and a direction of preferred orientation of that axis, the director. The brackets denote an averaging over all molecules under consideration. The orientation of the director may vary locally. However, it can be dictated by external fields or by surface anchoring conditions over macroscopic distances. Planar silica surfaces enforce planar anchoring of 7CB and 8CB without a preferred lateral orientation \cite{Kumar2001}. Additionally, the director is expected to be oriented parallel to the long axis in a cylindrical silica channel \cite{GrohDietrich1999}. A statement which shall be explored in the following by birefringence measurements.

The propagation speed of light and thus the refractive index $n$ in a LC sensitively depends on the orientation of the polarization with respect to the molecular orientation of the anisotropic nematogens. Conversely, the state of molecular order in a LC can be inferred from optical polarization measurements. To a good approximation, $Q$ is proportional to the optical birefringence $\Delta n=n_{\rm e}-n_{\rm o}$, where $n_{\rm o}$ and $n_{\rm e}$ refer to polarizations perpendicular and parallel to the local optical axis, the so-called ordinary and extraordinary refractive indices \cite{Haller}, respectively. In a nematic LC the local optical axis agrees with the director. Thus, in principle it is sufficient to determine the experimentally accessible $\Delta n$ in order to determine the molecular arrangement in an LC. However, there are weak, but final $T$-dependencies of the bare refractive indices, which do not originate from changes of the averaged collective molecular orientations but from changes in the anisotropic molecular polarizabilities of the single molecules as a function of $T$. In order to separate out these effects, we resort to the quantity $\Delta n^*=n_{\rm e}^2(T)-n_{\rm o}^2(T)\propto \Delta n(T) \cdot (n_{\rm e}(T)+n_{\rm o}(T))/2$ which can be shown to be solely proportional to $Q(T)$ \cite{Lau}. For simplicity, we will refer to $\Delta n^*$ as ''effective birefringence'' in the following.

For our measurements, a monolithic silica membrane permeated by an array of parallel aligned, non-interconnected channels of 300~$\mu$m length was prepared by thermal oxidation of a free-standing silicon membrane \cite{Gruener2008} at 800~\dC~for 12~hours. The mean channel diameter $D$ and porosity $P$ was determined by recording a volumetric nitrogen sorption isotherm at $T~=~77~K$ to $D$=10.0$\pm0.5$~nm and $P$ = 50$\pm$2\%, respectively. Electron micrographs of the channels \cite{Gruener2008} indicate sizeable 1.0$\pm0.5$~nm mean square deviations of their surfaces from an ideal cylindrical form. The membrane was completely filled with the LCs by spontaneous imbibition \cite{Huber2007}. For the bulk measurement, we used a 50~$\mu$m thick glass cell containing homeotropically aligned LCs, see Fig.~\ref{fig1}(a). 

\begin{figure}[htbp]
\epsfig{file=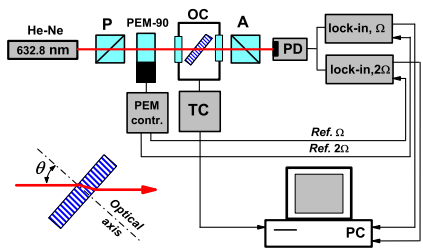, angle=0, width=0.7\columnwidth}\caption{(color online). Schematic experimental setup of a high-resolution birefringence
measurement consisting of a He-Ne laser, an optical polarizer (P), a temperature controlled (TC) optical cell (OC), an optical analyzer (A), a photoelastic modulator (PEM-90), a photodiode as a detector (PD) and a ''lock-in'' detection and analyzing unit. The inset depicts the orientation of the silicon nanochannel membrane with respect to the incident laser beam.\label{fig3}}
\end{figure}

We used a high-resolution optical modulated beam method for the accurate determination of the phase retardation $R$ between two perpendicularly polarized components of light transmitted through the samples. The setup, see Fig.~\ref{fig3}, employs an optical photoelastic modulator and a dual lock-in detection scheme in order to minimize the influence of uncontrolled light-intensity fluctuations \cite{Skarabot, Kumar2001}. After passing the sample the laser light intensity ($\lambda=632.8~nm$) was detected by a photodiode and two lock-in amplifiers, which simultaneously determined the amplitudes of the first ($U_{\rm \Omega}$) and second ($U_{\rm 2\Omega}$) harmonics, respectively. The phase retardation by the sample $R=\arctan[(U_{\rm \Omega}J_{\rm 2}(A_{\rm 0}))/(U_{\rm 2\Omega}J_{\rm 1}(A_{\rm 0}))]$ (here $J_{\rm 1}(A_{\rm 0})$ and $J_{\rm 2}(A_{\rm 0})$ are the Bessel functions corresponding to the PEM retardation amplitude $A_{\rm 0}=0.383\lambda$) was measured for an incident angle $\theta=43.5$ deg between laser beam and sample surface and thus between beam and long axes of the silica channels, see inset of Fig.~\ref{fig3}. For such a tilted sample geometry the conversion of the retardation $R$ to $\Delta n$ and $\Delta n^*$ was performed by numerically solving Berek's compensator formula \cite{Berek}.

In Fig.~\ref{fig1} $\Delta n^*$ is plotted for bulk 7CB and 8CB upon slow cooling ($\sim$0.01 K/min) to the solidification temperature. There is a jump in $\Delta n^*(T)$ of bulk 7CB typical of the first-order \textit{I-N} phase transition at $T^{\rm b}_{\rm IN}\approx$~42~\dC \cite{Lau}. $\Delta n^*(T)$ of 8CB exhibits the signatures characteristic of the first-order \textit{I-N} transition at $T^{\rm b}_{\rm IN}\approx$~41~\dC~and an almost continuous \textit{N-SmA} transition at $T^{\rm b}_{\rm NSmA}\approx$34~\dC \cite{I1,Kas}. Any pretransitional effects are clearly absent in the bulk isotropic phase of both LCs investigated, $\Delta n^*(T)=0$ for $T>T^{\rm b}_{\rm IN}$.

The nanoconfined LCs reveal a considerably different behavior, see Fig.~\ref{fig1} lower panels. In agreement with neutron diffraction experiments on 8CB in nanochannels, there is no indication of a sharp \textit{N-SmA} phase transition \cite{Guegon2006}. More interestingly, there exists a residual $\Delta n^*$ characteristic of a paranematic LC state at $T$s far above $T^{\rm b}_{\rm IN}$. Upon further cooling $\Delta n^*$ increases continuously and at the lowest $T$s investigated, the absolute magnitude of $\Delta n^*$ is compatible with an 80\% (75\%) alignment of the 7CBs' (8CBs') long axes parallel to the channel axes. Hence, the silica nanochannel confinement dictates indeed a substantial molecular alignment, as proposed in the introduction and illustrated in the inset of Fig.~\ref{fig1}. More importantly, it renders the transition continuous.

\begin{figure}[htbp]
 \epsfig{file=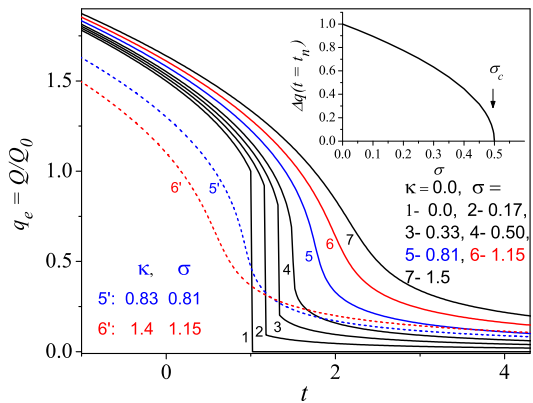, angle=0,
width=0.9\columnwidth}\caption{(color online). Order parameter $q_{\rm e}$ in the KKLZ-model as a function of reduced temperature $t$ for selected effective surface fields $\sigma$ in the absence of quenched disorder, $\kappa=0$ (solid lines) and for the $\sigma$-, $\kappa$-values which yield the best fits of $\Delta n^*(T)$ measured for the confined LCs (dashed lines). Inset: Order parameter jump at the \textit{I-N (P-N)} transition as a function of the surface field $\sigma$ ($\kappa=0$).}
\label{fig2}
\end{figure}

In the following we are going to analyze the peculiar $\Delta n^*$ behavior within a Landau-de~Gennes model for the $I-N$ transition in confinement suggested by Kutnjak, Kralj, Lahajnar, and Zumer (KKLZ-model) \cite{Kralj, Sheng1976}. The dimensionless free energy density of a nematic phase spatially confined in a cylindrical geometry with planar anchoring conditions reads in the KKLZ-model as:
\begin{equation}
f=tq^2-2q^3+q^4-q\sigma+\kappa q^2 \nonumber \label{eq2}
\end{equation}
where $q=Q/Q(T^{\rm b}_{\rm IN})$ is the scaled nematic order parameter, $t$ is a reduced temperature, and $\sigma$ is the effective surface field. The last term in Eq.~1 describes quenched disordering effects due to surface-induced deformations (wall irregularities) \cite{I1}. Minimalization of $f$ yields the equilibrium order parameter $q_{\rm e}$, which is shown in Fig.~\ref{fig2} for selected values of $\sigma$ and $\kappa$ as a function of $t$. In the KKLZ-model, the \textit{I-N} transition is of first order for $\sigma < \sigma_{\rm c}=0.5$. The jump of $q_{\rm e}$ approaches zero while $\sigma \rightarrow 0.5$, see inset in Fig.~\ref{fig2}. Thus, $\sigma_{\rm c}$ marks a critical threshold separating first-order, discontinous from continuous \textit{I-N} behavior.

In the following we apply the KKLZ-model to our measured $\Delta n^*(T)$. The solid lines in Figs.~\ref{fig1}(a),(b) are the best fits of the dependencies $\Delta n^*(T)$ as obtained by rescaling $q_{\rm e}$ and $t$ while assuming an absence of any surface ordering and quenched disorder fields in the bulk state ($\sigma(bulk)=\sigma(D=\infty)$=0, $\kappa=0$). Thereby, we achieve an encouraging agreement between the measured $\Delta n^*(T)$ curves in the proximity of $T^{\rm b}_{\rm IN}$ and deep into the nematic phase for both bulk LCs. The sizeable deviations for 7CB below 230~\dC~originate in the neglect of higher order terms in the KKLZ-model, necessary to produce the saturation behavior of $q$ at low $T$.

More importantly, we achieve also an excellent agreement between measured and KKLZ-modelled $\Delta n^*(T)$ over the entire \textit{P-N} transition regime for the nanoconfined LCs, provided we assume final surface ordering fields of magnitude $\sigma(D=10$nm$)$ = 0.81, 1.15 and strengths of the quenched disorder parameter $\kappa$ = 0.83, 1.4 for 7CB and 8CB, resp., see panel (c) and (d) in Fig.~\ref{fig1}, respectively. As expected from the observed continuous behavior of $\Delta n^*(T)$ in both cases $\sigma(D)$ is well above $\sigma_{\rm c}=0.5$. Moreover, the upward shift in $T_{\rm PN}$ due to the anchoring surface field, predicted by the KKLZ-model, is slightly overbalanced by a downward shift due to quenched disorder effects for this set of KKLZ-parameters - see also Fig. \ref{fig2}.

The bare geometrical mechanism which favors an alignment of rod-like molecules in cylindrical confining geometries is expected to increase in strength with the length of the molecules \cite{GrohDietrich1999}. Also the alignment effect of the anchoring field of the confining walls increases with this length. Therefore the $\sigma$-increase of $30\%$ appears not too surprising, despite the relatively small 10\% length-increase between 8CB and 7CB.

According to the KKLZ-model $\sigma(D)$ scales with $1/D$, which also allows us to estimate $\sigma$ as a function of $D$ for other confining silica geometries. In particular, we arrive at a $D_{\rm c}$ of 23~nm and 16~nm for the ''critical'' silica pore diameter of 8CB and 7CB, resp., ($\sigma(D_{\rm c})=\sigma_{\rm c}=0.5$) separating continuous from discontinuous behavior. This is in quantitative agreement with the still weakly discontinuous behavior reported for 8CB confined in 24~nm xerogel mesopores \cite{Kralj} and the continuous behavior found for 7CB in 7~nm mean pore diameter Vycor glass \cite{I1}.


In conclusion, our measurements indicate that in $10$~nm straight silica channels the surface anchoring fields render the bulk discontinuous \textit{I-N} transition to a continuous \textit{P-N} transition. The transition temperature is changed only marginally, due to a balance of its surface ordering induced upward and its quenched disorder induced downward shift, similarly as has been observed for LCs imbibed in tortuous Vycor glass pores and a variety of other mesoporous matrices \cite{I1}. The purely phenomenological findings presented here would profit from more microscopic information, for example gained by Monte Carlo or Molecular Dynamics simulations \cite{Steuer2005} or from x-ray and neutron scattering experiments.

Finally we would like to state, that the rheology of bulk LC changes abruptly at the \textit{I-N} transition \cite{Miesowicz1946, Jadzyn2001}. By contrast, the gradual \textit{P-N} transition reported here should lead to a continuous $T$-evolution of the fluidity of LC in silica nanochannels. Moreover, velocity slippage at the walls, crucial in the emerging field of nanofluidics \cite{MNFluidics, Huber2007}, is expected to be associated with the molecular alignments in the channels \cite{Heidenreich2007}. Thus, we hope the peculiarities reported here will stimulate LC flow experiments with silica nanochannels, which would be important extensions of previous experiments on the fluidity of LCs in thin film geometries \cite{RheologyThinFilms}.

\begin{acknowledgments}
We thank the DFG for support within the priority program 1164, \textit{Nano- \& Microfluidics} (Hu 850/2) and acknowledge support by the DAAD and the French Ministry of Foreign Affairs within the French-German \mbox{PROCOPE} program.
\end{acknowledgments}

\end{document}